\documentclass[12pt]{iopart}
\usepackage{graphicx} 
\usepackage{tabularx}
\usepackage{array}
\expandafter\let\csname equation*\endcsname\relax
\expandafter\let\csname endequation*\endcsname\relax
\usepackage{amsmath}
\usepackage{subcaption}
\newcommand\bea{\begin{eqnarray}}
\newcommand\eea{\end{eqnarray}}
\newcommand\be{\begin{equation}}
\newcommand\ee{\end{equation}}
\newcommand\ba{\begin{align}}
\newcommand\ea{\end{align}}

\DeclareUnicodeCharacter{2212}{\textendash}
\begin{document}

\title{Impact of Cosmology on Lorentz Invariance Violation Constraints from GRB Time-Delays}
\author{Denitsa Staicova }
\address{Institute for Nuclear Research and Nuclear Energy, Bulgarian Academy of Sciences, Sofia, Bulgaria}
\ead{dstaicova@inrne.bas.bg}
\begin{abstract}
Putting constraints on a possible Lorentz Invariance Violation (LIV) from astrophysical sources such as gamma-ray bursts (GRBs) is essential for finding evidences of new theories of quantum gravity (QG) that predict an energy-dependent speed of light. This search has its own difficulties, so usually, the effect of the cosmological model is understudied, with the default model being a fixed-parameters $\Lambda$CDM.  In this work, we use various astrophysical datasets to study the effect of a number of dark energy models on LIV constraints. To this end, we combine two public time-delay GRB  datasets with the supernovae Pantheon dataset, several measurements of angular baryonic acoustic oscillations (BAO), the cosmic microwave background (CMB) distance prior and an optional GRB or quasars dataset. For the LIV parameter $\alpha$, we find the expected from previous works average value of $\alpha \sim 4 \times 10^{-4}$, corresponding to $E_{QG}\ge 10^{17}$ GeV for both time-delay (TD) datasets, with the second one being more sensitive to the cosmological model. The cosmology results in a minimum 20\% deviation in our constraints on the energy. Interestingly, adding the TD points makes the DE models less-preferable statistically and shifts the value of the parameter $c/(H_0 r_d)$ down, making it smaller than the expected value. We observe that possible LIV measurements critically depend on the transparency of the assumptions behind the published data concerning cosmology, and taking this into account may be an important contribution in the case of possible detection.
\end{abstract}

\date{March 2023}

\maketitle

\section{Introduction}

Current astronomical surveys have provided us with an abundance  of new data to tackle important cosmological problems such as nature of dark energy and dark matter, the Hubble tension and others \cite{Abdalla:2022yfr}. They can, however, be applied to another very important question -- the quest for finding a theory of quantum gravity (QG) and more specifically, for probing for Lorentz Invariance Violation (LIV). LIV measures the violation of relativistic symmetries caused by the dispersion in vacuum of different messengers --  photons, neutrinos and gravitational waves. Several theories, such as string theory, Horava–Lifshitz gravit, emergent gravity etc \cite{Addazi:2021xuf} which attempt to unify the force of gravity with the other three fundamental forces of matter, predict LIV. Detecting evidences of LIV would provide insights into the possible existence of a QG theory. Given that LIV is expected to be amplified over cosmological distances, using high-redshift high-energy astronomical objects is key to trying to measure the small Planckian effect (at $E<<E_{Pl})$. It is worth noting that studies on LIV effects have a long history in the field \cite{Colladay:1996iz,  AmelinoCamelia:1997gz, Colladay:1998fq,Kostelecky:1988zi, Ellis:2005sjy, Jacob:2008bw, Gubitosi:2009eu, Vasileiou:2015wja, Amelino-Camelia:2013wha, Magueijo:2001cr, Amelino-Camelia:2000bxx, Amelino-Camelia:2000stu, Amelino-Camelia:2002cqb, Kostelecky:2008ts}, for reviews on the topic, see \cite{Wei:2021ite, Wei:2021vvn, Zhou:2021ycu, Addazi:2021xuf, Desai:2023rkd}. 

A perfect probe for LIV effects are gamma-ray bursts (GRBs). GRBs are among the most powerful events in the Universe, with isotropic energy $E_{iso} \ge 10^{52}$ erg (for example GRB 221009A is with $\sim 1.2 \times 10^{55}$ erg \cite{Burns:2023oxn}, see within for a table of GRBs with $E_{iso} \ge 10^{54}$ erg). Some have been observed at redshift up to $z\sim 9.4$ (in the case of GRB 090429B \cite{Cucchiara:2011pj}). Moreover, their emission have been detected in the highest energy bands, for example GRB 221009A exhibited very high emission at energies above 10 TeV (\cite{huang2022lhaaso,HESS:2023qhy}).  The combination of high energy emission and large distance offers a unique opportunity to measure LIV. The results so far have been summarized in \cite{Addazi:2021xuf} and they depend on the messenger in use: for photons from GRBs,  one finds as an upper bound for possible LIV:
$E_{QG,1} \ge 2.23 \times 10^{14}$ GeV and $E_{QG,2}\ge 0.87 \times 10^{6} $ GeV from the 0.2TEV GRB 190114C \cite{Du:2020uev} (or the compatible with no time delay, $E_{QG,1} \ge 0.58 \times 10^{19}$ GeV and $E_{QG,2} \ge 6.2 \times 10^{10}$ GeV according to \cite{MAGIC:2020egb})
and  $E_{QG,1} > 9.3 \times10^{19}$ GeV and $E_{QG,2}>1.3 \times 10^{11}$ GeV from GRB 090510 \cite{Vasileiou:2013vra}
\footnote{The LIV constraints are usually given in GeV units since the characteristic scale of QG is considered to be the Planck scale $E_{Pl}\approx 1.22\times 10^{19}$ GeV}. In the model with an energy-dependent intrinsic time-delay introduced in \cite{Wei:2016exb}, one gets $E_{QG,1} \ge 0.3 \times 10^{15}$ GeV from GRB 160625B and 23 more GRBs \cite{Pan:2020zbl}, which has been further extended in \cite{Agrawal:2021cim} to obtain $E_{QG}^1\ge 10^{16}$ GeV. 

There are some important considerations when it comes to using GRBs for constraining LIV. The observed time-delay is a combination between the time-delay at the source (intrinsic to the GRB), the time-delay due to QG and other factors, primarily related to the propagation of the EM waves. These propagation effects include dispersion by the line-of-
sight free electron content, non-zero rest mass of the messenger, and the Shapiro effect \cite{Shapiro:1964uw, Longo:1987gc, Minazzoli:2019ugi, Laanemets:2022rmn}. For high energy photons, most propagation effects can be ignored, as discussed in \cite{Addazi:2021xuf}. The time-delay at the source, however, depends on the model of the GRB progenitor. In general, GRBs fall into two groups -- long GRBs produced by the core-collapse of a massive star, and short GRBs, produced by neutron star and/or black hole mergers. These two classes suggest different emission mechanisms, which need to be taken into account when attempting to measure LIV. Moreover some GRBs cannot be clearly classified, adding further complexity to the problem \cite{Bernardini:2014oya}. For more details on the methods to handle this issue, see \cite{Addazi:2021xuf}. Finally, there is the question of the underlying cosmology. Since the redshifts at which GRBs are observed are high, the cosmological model can affect the observed GRB quantities. In general, one assumes $\Lambda$CDM since it is the concordance model in cosmology, but as summarized in \cite{Abdalla:2022yfr}, the tensions observed in the Hubble constant and other quantities leave the door open for other alternative models. First attempts to take into account different cosmological models in LIV studies have been made in \cite{Biesiada:2007zzb, Biesiada:2009zz, Zou:2017ksd} by incorporating EOS such as quintessence and Chaplygin gas, as well as the cosmography approach. These studies found that the uncertainty in these models significantly affects the final LIV constraints. 

In this paper, we utilize two available GRB time-delay (TD) datasets and we combine them with several robust datasets used extensively in the literature. These datasets include  measurements of baryonic acoustic oscillations (BAO), supernovae type IA, GRBs, quasars and the CMB distance prior.  Our goal is to investigate how different dark energy (DE) models and a model with spatial curvature $\Omega_K$CDM affect the final constraints in two aspects -- with respect to the dark energy parameters and with respect to the LIV parameters. We show that different cosmological models affect the LIV results differently for the two TD datasets, possibly due to differences in their processing methods. We also see that the TD dataset influences the cosmological parameters yielding interesting results on the quantity $c/H_0 r_d$, which appears lower than expected. The paper is organized as follows: in Section \ref{sec:theory}, we discuss the Theoretical background, in Section \ref{sec:methods} we describe our Methods, in Section \ref{sec:datasets} we elaborate on our Datasets, in Section \ref{sec:results} we discuss the numerical results and in Section \ref{sec:discussion}, one can find a Discussion on the obtained results. 

\section{Theoretical background}
\label{sec:theory}
\subsection{Definition of LIV quantities}
The modified dispersion relation for photons ~\cite{AmelinoCamelia:1997gz} can be approximated on low energy scales ($E\ll E_{QG}$) (see ~\cite{Vasileiou:2013vra}) to,
 \be
E^2=p^2 c^2\left[1-s_\pm\left(
 \frac{E}{\xi_n E_{QG}}\right)^n\,\right]\,,
 \ee
 where 
$E_{QG}$ is the effective QG energy scale,  $c$ is the speed of light, $p$ and $E$ are the momentum and energy of photons, respectively. Here $s_\pm=\pm 1$ for subluminal resp. superluminal propagation ~\cite{Vasileiou:2013vra}. $\xi_n$ is a dimensionless
 parameter, and $E_{QG,n}=\xi_n E_{QG}$ is  the
 effective energy scale of order $n$ at which LIV happens  ~\cite{Vasileiou:2013vra},  where in this work, we consider only $n=1$. This is one of the possible anzatzes for a modified dispersion relation, see \cite{Amelino-Camelia:2000stu, Liberati:2004ju, Jacob:2008bw, Zloshchastiev:2009zw, Bezerra:2019vrz, Furtado:2021aod}, which we prefer due to its prevalence in the literature enabling us to easily compare our results to earlier ones.  
 
Following ~\cite{Jacob:2008bw} and \cite{Biesiada:2007zzb, Biesiada:2009zz}, the time of flight of the photon of energy $E$ for a redshift $z$ is equal to 
\be
t = \int_0^z [1 + \frac{E}{E_{QG}}(1 + z')] \frac{dz'}{H(z')}
\label{eq:time-delay}
\ee
Consequently, the time delay between a low energy and a high energy photon with the
energy difference $\Delta E= E_{high}-E_{low}$ takes the following form for the subluminal case:
\be
\Delta t_{LIV} = \frac{\Delta E}{E_{QG}}\int_0^z (1 + z' )\frac{dz'}{H(z')}
\ee

\noindent where: $H(z) = H_0 E(z)$ is the Hubble parameter, with $H_0$ -- the Hubble constant at $z=0$ and $E(z)$ --  the equation of state of the Universe. This equation can be also obtained from general perturbation of the general relativistic dispersion relation of freely falling particles on homogeneous and isotropic spacetimes  \cite{Pfeifer:2018pty}.

As mentioned in the Introduction, for GRBs, the observed time
 delay between two different energy bands depends not only on the LIV time-delay, but also on more terms ~\cite{Wei:2015hwd,Gao:2015lca}. The most important among them is the intrinsic time delay, $\Delta t_{\rm int}$, related to the physics of the source of the GRB (i.e. the photons were not emitted simultaneously due to the processes in the GRB's central engine). Since we cannot predict that term without a very good knowledge of the source, we can quantify it as (e.g.~\cite{Ellis:2005sjy,Biesiada:2009zz,Pan:2015cqa}),
 $\Delta t_{\rm int}=\beta\left(1+z\right)$, where $\beta$ will be a free parameter in our model. All the other terms are negligible according to our current knowledge -- meaning that we consider no photon mass at rest, no contribution from the dispersion by the
 line-of-sight free electron content for GRB photons and no contribution from gravitational potentials along the
 propagation path of photons, for more details see \cite{Zou:2017ksd}).
 
 In this case, our final equation for the GRB time-delay becomes: 
 \be
 \frac{\Delta t_{obs}}{1+z}=a_{LIV}K+\beta\,,
 \ee
 where $a_{LIV}\equiv\Delta E/(H_0 E_{QG})$, and
 \be
 K\equiv\frac{1}{1+z}
 \int_0^z\frac{(1+\tilde{z})\,d\tilde{z}}{h(\tilde{z})}\,.
 \ee
 if $a_{LIV}=0$, there is no LIV, while if $a_{LIV}\not=0$, there is LIV on energy scales above $E_{QG}$. We see this term connects the LIV time-delays with the cosmological model hidden in $h(z)$. 

In our models we use the form: 
 \be
 \Delta t_{obs} = a_{LIV}{\cal K} + \beta\left(1+z\right)\,,
 \label{eq:tobs}
 \ee
where ${\cal K}=(1+z)K$. Note, below we denote $a_{LIV}=\alpha$ and the original parameter $b$ is replaced with $\beta$, to avoid mixing this parameter with $b=c/(H_0 r_d)$.

\subsection{Cosmology}
We assume the  Friedmann - Lema\^itre - Robertson - Walker metric with the scale parameter $a = 1/(1+z)$. The equation of state of the Universe for $z-$dependent DE then is:
\begin{equation}
    E(z)^2 = \Omega_{r} (1+z)^4 + \Omega_{m} (1+z)^3 + \Omega_{K} (1+z)^2 + \Omega_{DE}(z),
    \label{eq:hz}
\end{equation}

\noindent where in standard $\Lambda$CDM, $\Omega_{DE}(z)\to \Omega_\Lambda$. The  expansion of the universe is given by $E(z)= H(z)/H_0$, where $H(z) := \dot{a}/a$ is the Hubble parameter at redshift $z$ and $H_0$ is the Hubble parameter today. $\Omega_{r}$, $\Omega_{m}$, $\Omega_{DE}$ and $\Omega_{K}$ are the fractional densities of radiation, matter, dark energy and the spatial curvature at redshift $z=0$. We set the radiation energy density to zero, i.e. $\Omega_r = 1 - \Omega_m - \Omega_{\Lambda} - \Omega_{K}=0$. The spatial curvature for the DE models is also zero, $\Omega_K=0$,  i.e. a flat Universe.

We will consider a number of different DE models, all of which will feature a dark energy component depending on $z$: Chevallier-Polarski-Linder (CPL, \cite{Chevallier:2000qy,Linder:2005ne,Barger:2005sb}), Barboza-Alcaniz (BA  \cite{Barboza:2008rh, Escamilla-Rivera:2021boq}),  Low correlation model (LC \cite{Wang:2008zh,  Escamilla-Rivera:2021boq}), Jassal-Bagla-Padmanabhan (JBP) parametrization \cite{Jassal:2004ej, Motta:2021hvl}, Feng--Shen--Li--Li parametrization~\cite{Feng:2012gf, Motta:2021hvl}.  As an alternative to $\Lambda$CDM, we consider the phenomenological Emergent Dark Energy (pEDE) model \cite{Li:2019yem, Li:2020ybr}. One can find the relevant equations for $\Omega_{DE}(z)$ to be substituted in Eq. \ref{eq:hz} in Table \ref{table:DE_models}. All of the equations of state recover $\Lambda$CDM for $w_0=-1, w_a=0$, except for pEDE which is an alternative to  $\Lambda$CDM. 

\begin{table}
	\begin{center}
		\begin{tabular}{|c|c|c|}
			\hline
			  Model & $\Omega_{DE}(z)= \Omega_{\Lambda} \times$&  $w(z)$  \\
   \hline
             CPL  &  $  \exp\left[\int_0^{z} \frac{3(1+w(z')) dz'}{1+z'}\right] $ & $w_0 + w_a \frac{z}{z+1}$  
             \\
             BA & $(1+z)^{3(1+w_0)}{(1+z^2)}^{\frac{3w_1}{2}}$ & $w_0+z\frac{1+z}{1+z^2}w_1$   \\
             LC & $
(1+z)^{(3(1-2w_0+3wa))} e^{\frac{9(w_0-wa)z}{(1+z))}}$ &  $\frac{(-z+z_c)w_0+z(1+z_c)w_c}{(1+z)z_c}$ \\
	JPB  & $(1+z)^{3(1+w_0)}e^{\frac{3w_1z^2}{2(1+z)^2}}$ & $ w_0+w_1\frac{z}{(1+z)^2}$ \\
  FSLLI & $ (1+z)^{3(1+w_0)}e^{ \frac{3w_1}{2}\arctan(z)}(1+z^2)^{\frac{3w_1}{4}}(1+z)^{- \frac{3}{2}w_1}$ & $w_0+w_1\frac{z}{1+z^2} $ \\
  FSLLII  & $(1+z)^{3(1+w_0)}e^{- \frac{3w_1}{2}\arctan(z)}(1+z^2)^{\frac{3w_1}{4}}(1+z)^{+ \frac{3}{2}w_1} $ & $w_0+w_1\frac{z^2}{1+z^2} $ \\
 PEDE  & $ \frac{1-\tanh(\bar{\Delta} \log_{10}(\frac{1+z}{1+z_t}))}{1+\tanh(\bar{\Delta} \log_{10}({1+z_t})}$ & $ -\frac{\left({1+{\rm{tanh}}\left[{\rm{log}}_{10}\,(1+z)\right]}\right)}{3 {\rm{ln}}\, 10}\! -\!1 $  \\
 
   \hline
		\end{tabular}
	\end{center}
	\caption{{ The DE models we use in this work. The references can be found in the text.}}
\label{table:DE_models}
\end{table}

In order to use our selected datasets, we need to introduce some quantities. 

The BAO measurements are defined trough the angular diameter distance,  $D_\textrm{A}$:

\be
D_\textrm{A}
=\frac{c}{(1+z) H_0 \sqrt{|\Omega_{K}|}  } \textrm{sinn}\left[|\Omega_{K}|^{1/2}\int_0^z \frac {dz'} {E(z')}\right]\ ,
\label{eq:DA}
\ee

where $\textrm{sinn}(x) \equiv \textrm{sin}(x)$, $x$, $\textrm{sinh}(x)$ for $\Omega_{K}<0$, $\Omega_{K}=0$, $\Omega_{K}>0$ respectively. We see that for the measured $D_A/r_d$ one can isolate the variable $b=c/(H_0 r_d)$. We do not use the radial BAO measurements in this work. 

The SN, the GRB and the Quasars datasets measure the distance modulus $\mu(z)$ which is related to the luminosity distance ($d_L = D_A(1+z)^2$) through:
\begin{equation}
        \mu_B (z) - M_B = 5 \log_{10} \left[ d_L(z)\right] + 25  \,,
\label{eq:dist_mod_def}
\end{equation}
where $d_L$ is measured in units of Mpc, and $M_B$ is the absolute magnitude.

To take into account the CMB, we are going to use the CMB distance priors. They provide information of CMB power spectrum in two aspects: the acoustic scale $l_\textrm{A}$ characterizes the CMB temperature power spectrum in the transverse direction and the "shift parameter" $R$ describes the CMB temperature spectrum along the line-of-sight direction \cite{Komatsu:2008hk}:
\begin{align*}
 l_\textrm{A} =(1+z_*)\frac{\pi D_\textrm{A}(z_*)}{r_s(z_*)} ,\\
R\equiv(1+z_*)\frac{D_\textrm{A}(z_*) \sqrt{\Omega_m } H_0}{c},
\label{la:Rz}
\end{align*}
where $z_* \approx 1089$ is the redshift at the photon decoupling epoch \cite{Aghanim:2018eyx}. $r_s(z_*)$ is the co-moving sound horizon at redshift $z_*$. In Ref. \cite{Chen:2018dbv}, the authors derive the distance priors in several different models using $Planck$ 2018 TT,TE,EE $+$ lowE latest CMB data \cite{Aghanim:2018eyx}. We use the provided in \cite{Chen:2018dbv} correlation matrices to obtain the covariance matrices for $l_A$ and $R$ for each model.

\section{Methods}
\label{sec:methods}
The datasets we use are chosen to provide robust information of the background cosmology needed to evaluate both the equation of state and the time-delay, Eq. \ref{eq:time-delay}. For the BAO dataset, the definition of the $\chi^2$ which we will minimize is the standard one, since we use only uncorrelated angular measurement and we do not use covariance matrix for it.
\begin{equation}
\chi^2_{BAO} = \sum_{i} \frac{\left(\vec{v}_{obs} - \vec{v}_{model}\right)^2}{\sigma^2}.
\label{eq:chi_bao}
\end{equation}
Here $\vec{v}_{obs}$ is a vector of the observed points, $\vec{v}_{model}$ is the theoretical prediction of the model and $\sigma$ is the error of each measurement. To avoid setting priors on $H_0$ and $r_d$, we consider the quantity $b=c/(H_0 r_d)$, this way we do not calibrate our model with the early or late universe. 

Fot the SN and the GRB datasets, we use the approach used in \cite{Staicova:2021ntm} allowing us to marginalize over $H_0$ and $M_B$, so that we avoid setting priors on $H_0$ and $M_B$. The integrated $\chi^2$ in this case is (\cite{DiPietro:2002cz,Nesseris:2004wj,Perivolaropoulos:2004yr,Lazkoz:2005sp}):
\begin{equation}
\tilde{\chi}^2_{SN,  GRB} = D-\frac{E^2}{F} + \ln\frac{F}{2\pi},
\end{equation}
for
\begin{subequations}
\begin{equation}
D = \sum_i \left( \Delta\mu \, C^{-1}_{cov} \, \Delta\mu^T \right)^2,
\end{equation}
\begin{equation}
E = \sum_i \left( \Delta\mu \, C^{-1}_{cov} \, E \right),
\end{equation}
\begin{equation}
F = \sum_i  C^{-1}_{cov}  ,
\end{equation}
\end{subequations}
where  $\mu_{}^{i}$ is the observed luminosity, $\sigma_i$ is its error, and the $d_L(z)$ is the luminosity distance, $\Delta\mu =\mu_{}^{i} - 5 \log_{10}\left[d_L(z_i)\right)$, $E$ is the unit matrix, and $C^{-1}_{cov}$ is the inverse covariance matrix of the dataset. For the GRB dataset, $C^{-1}\to 1/\sigma_i^2$ since there is no known covariance matrix for it.  For the Pantheon dataset, we consider the total covariance  $C_{cov}=D_{stat}+C_{sys}$, where $D_{stat}=\sigma_i^2$ comes from the measurement and $C_{sys}$ is provided separately \cite{Deng:2018jrp}. We do not use the newer Pantheon+ dataset \cite{Scolnic:2021amr, Brout:2022vxf} , since at the time of writing, its covariance matrix is not yet public. 

Additionally, in order to avoid the effect of double-counting certain GRBs that are part of both the GRB dataset and the TD datasets, we use the quasars dataset for which the  $\chi^2$ is defined the same way as for GRBs and SNs. 

Finally, we define the time-delay likelihood which is the same as Eq. \ref{eq:chi_bao} but here the quantity we consider is the theoretical time-delay ($\vec{\Delta t}_{model}$) as defined in Eq. \ref{eq:tobs} and its observational value (($\vec{\Delta t}_{obs}$) provided by the TD dataset. 
\begin{equation}
\chi^2_{TD} = \sum_{i} \frac{\left(\vec{\Delta t}_{obs} - \vec{\Delta t}_{model}\right)^2}{\sigma^2},
\label{eq:chi_TD}
\end{equation}

The final $\chi^2$ is:
$$\chi^2=\chi^2_{BAO}+\chi^2_{CMB}+\chi^2_{SN}+\chi^2_{TD}(+ \chi^2_{GRB})(+ \chi^2_{Qua}).$$

\section{Datasets}
\label{sec:datasets}
The BAO dataset we are using is a collection of points from different  observations \cite{Chuang:2016uuz,Alam:2016hwk, Beutler:2016ixs,Blake:2012pj,Carvalho:2015ica,Seo:2012xy,Sridhar:2020czy,Abbott:2017wcz,Tamone:2020qrl,Zhu:2018edv,Hou:2020rse,Blomqvist:2019rah,Bourboux:2017cbm} selected not to be correlated by the approach described by \cite{Kazantzidis:2018rnb,Benisty:2020otr}. The CMB distant prior is given by \cite{Chen:2018dbv}. The SN data comes from the binned Pantheon dataset, which contains $1048$ supernovae luminosity measurements in the redshift range $z\in (0.01,2.3)$ \cite{Pan-STARRS1:2017jku} binned into 40 points. The GRB dataset \cite{Demianski:2016zxi} consists of 162 measurements in the range $z\in [0.034,9.3]$. The quasars dataset consists of 24 points in the range $z=0.079-5.93$ \cite{Roberts:2017nkm}.  We use it as an alternative to GRB dataset, even though both datasets can have unknown correlation or biases \cite{Dainotti:2022rfz}. 

To study the time delays, we use two different time delays (TD) datasets -- TD1 which was given by \cite{Ellis:2005sjy} and TD2 provided by \cite{Vardanyan:2022ujc}. TD1 consists of 9 BATSE GRBs (time resolution of 64ms in 4 channels in the range 25-320kEV), 15 HETE bursts (with resolution of 164ms, in similar range) and 11 SWIFT bursts (with resolution of 64ms in 4 channels in similar range) gathered from 1999 to 2005. The time lags have been obtained with the wavelet method which identifies unique light curve features from which it derives the  timelags by comparing the 115-320 keV band to the 25-55 keV one and renormalising to make the data compatible. The redshift range for TD1 is $z\in [0.168, 6.26]$. It has been used already in the works by \cite{Biesiada:2007zzb, Biesiada:2009zz} to describe the uncertainty brought into the constraints by variations in the cosmology model.  TD2 uses combined sample of 49 long and short GRBs observed by Swift dating between 2005 to 2013. In this dataset \cite{Bernardini:2014oya}, the time lags have been extracted trough a discrete cross-correlation function (CCF) analysis  between characteristic rest-frame energy bands of 100–150 keV and 200–250 keV.  The redshift for TD2 is $z\in [0.35, 5.47]$. We do not use newer results on time delays from more recent GRBs because they focus on the TD in different bands of a single GRB which makes the result more prone to deviations due to intrinsic effects \cite{Ellis:2005sjy}. 

Another important question is about properly extending our analysis to higher redshifts which are not accessible to BAO and SN. To this end, we have two choices -- the GRB dataset and the Quasars dataset. On one hand, the GRB extend to higher redshifts, but on the other, there is the problem of double-counting some of the GRBs which repeat in both the GRB and the TD datasets (about 25 GRBs). Even though the two datasets measure different things -- distance modulus vs. time delays, we do not know the correlation between these quantities. For this reason, we perform each of our numerical experiments in 3 variants -- the TD+SN+BAO datasets alone (labeled below $TD$), the former combined with the GRB dataset (labeled $TD+GRB$)  and in addition, we replace the GRB with the quasars dataset (called $TD+Qu$). The TD dataset by itself already includes high-redshift entries and the cosmological model enters trough Eq.\ref{eq:tobs}, but having more such points is important for the DE models since they depend on the redshift. 

To run the inference, we use a nested sampler to find the best fit. We use the open-source package $Polychord$ \cite{Handley:2015fda} with the $GetDist$ package \cite{Lewis:2019xzd} to present the results. 

The prior is a uniform distribution for all the quantities: $\Omega_{m} \in [0., 1.]$, $\Omega_{\Lambda}\in[0., 1 - \Omega_{m}]$, $c/ (H_0 r_d) \in [25, 35]$, $w_0 \in [-1.5, -0.5]$ and $w_a \in [-0.5, 0.5]$. Since the distance prior is defined at the decoupling epoch ($z_*$) and the BAO -- at drag epoch ($z_d$), we parametrize the difference between $r_s(z_*)$ and $r_s(z_d)$ as $rat= r_*/r_d$, where the prior for the ratio is $rat \in [0.9, 1.1]$, $\Omega_K \in [-0.1,0,1]$ .

\section{Results}
\label{sec:results}
The results we obtain for the different datasets are presented below and in the tables in the Appendix: Tables \ref{tab:allTD1}, \ref{tab:allTD2}, \ref{tab:statTD1}, \ref{tab:statTD2}. 

\begin{figure*}
 	\centering
\includegraphics[width=0.49\textwidth]{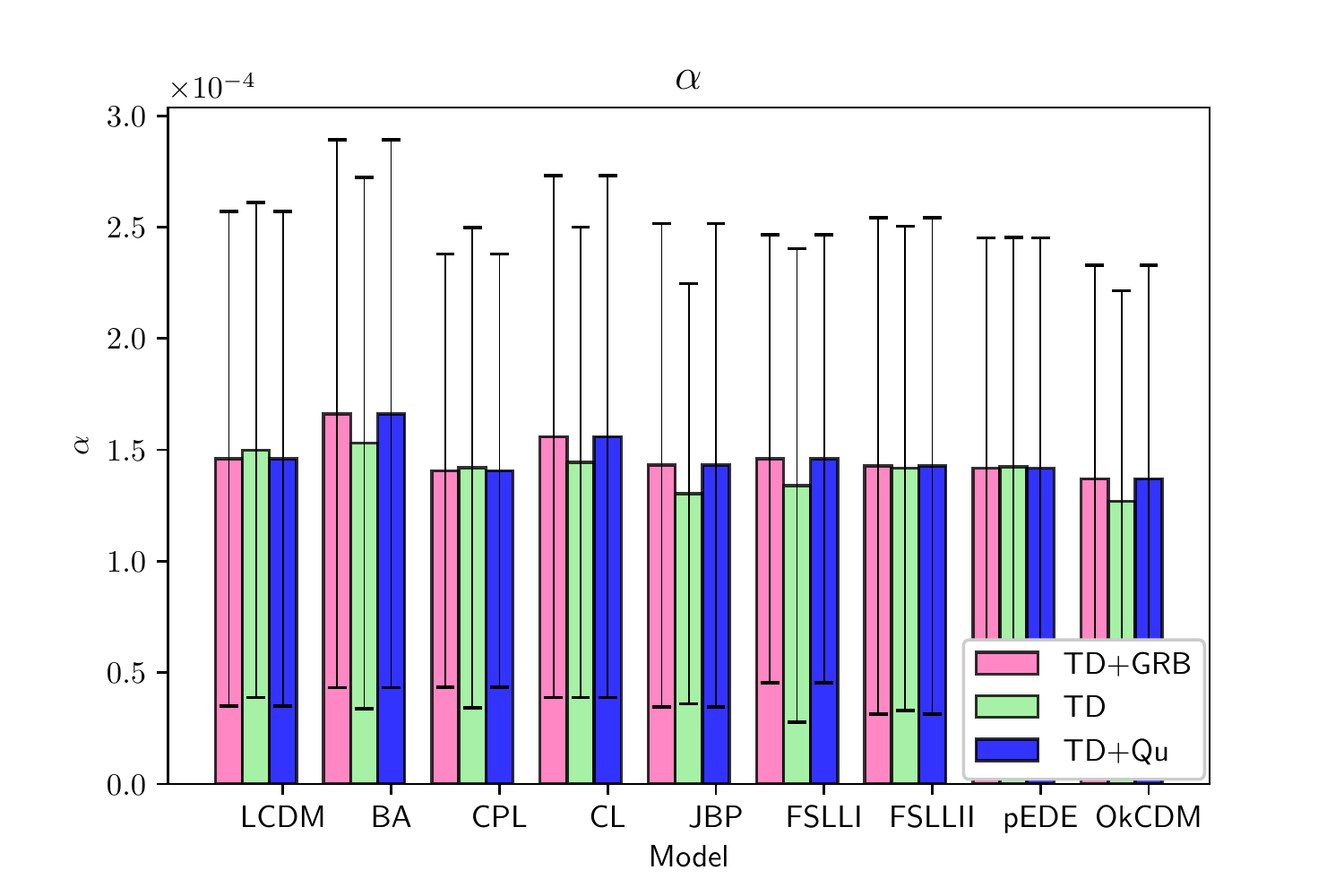}
\includegraphics[width=0.49\textwidth]{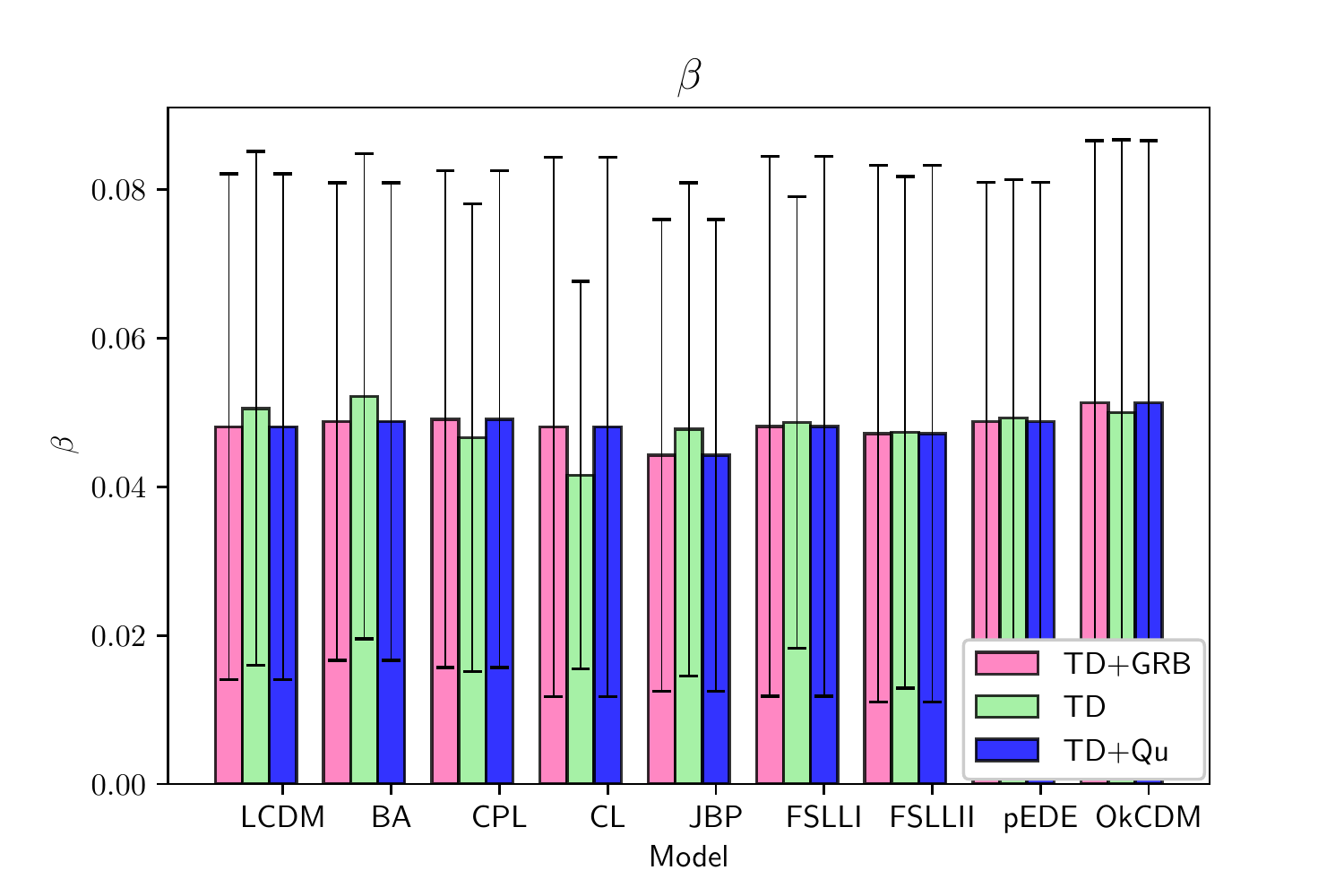}
\caption{The inferred values of the mean and error for the TD parameters $\alpha$ and $\beta$ for the first dataset, TD1 }
\label{fig1}
\end{figure*}

\begin{figure*}
 	\centering
\includegraphics[width=0.49\textwidth]{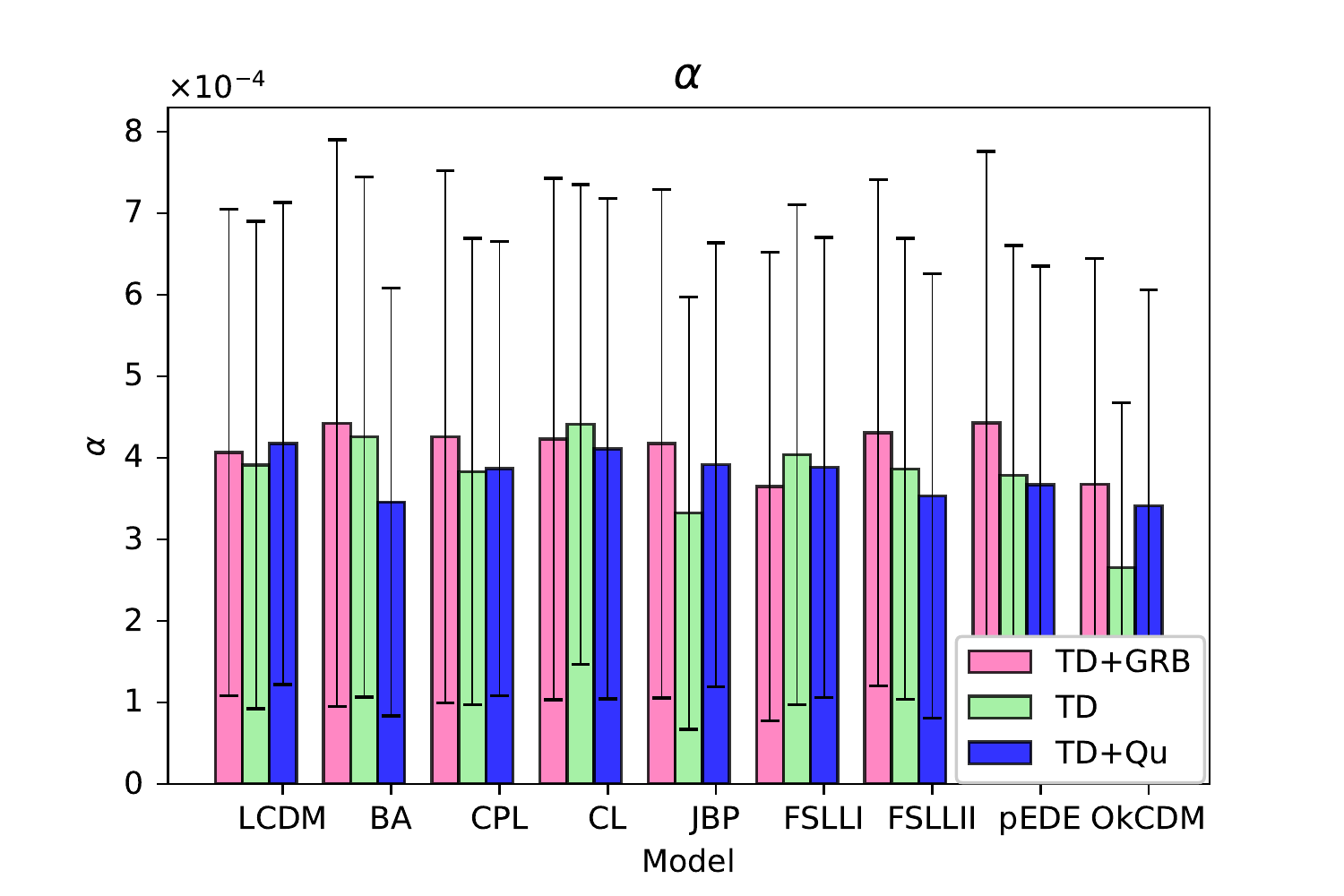}
\includegraphics[width=0.49\textwidth]{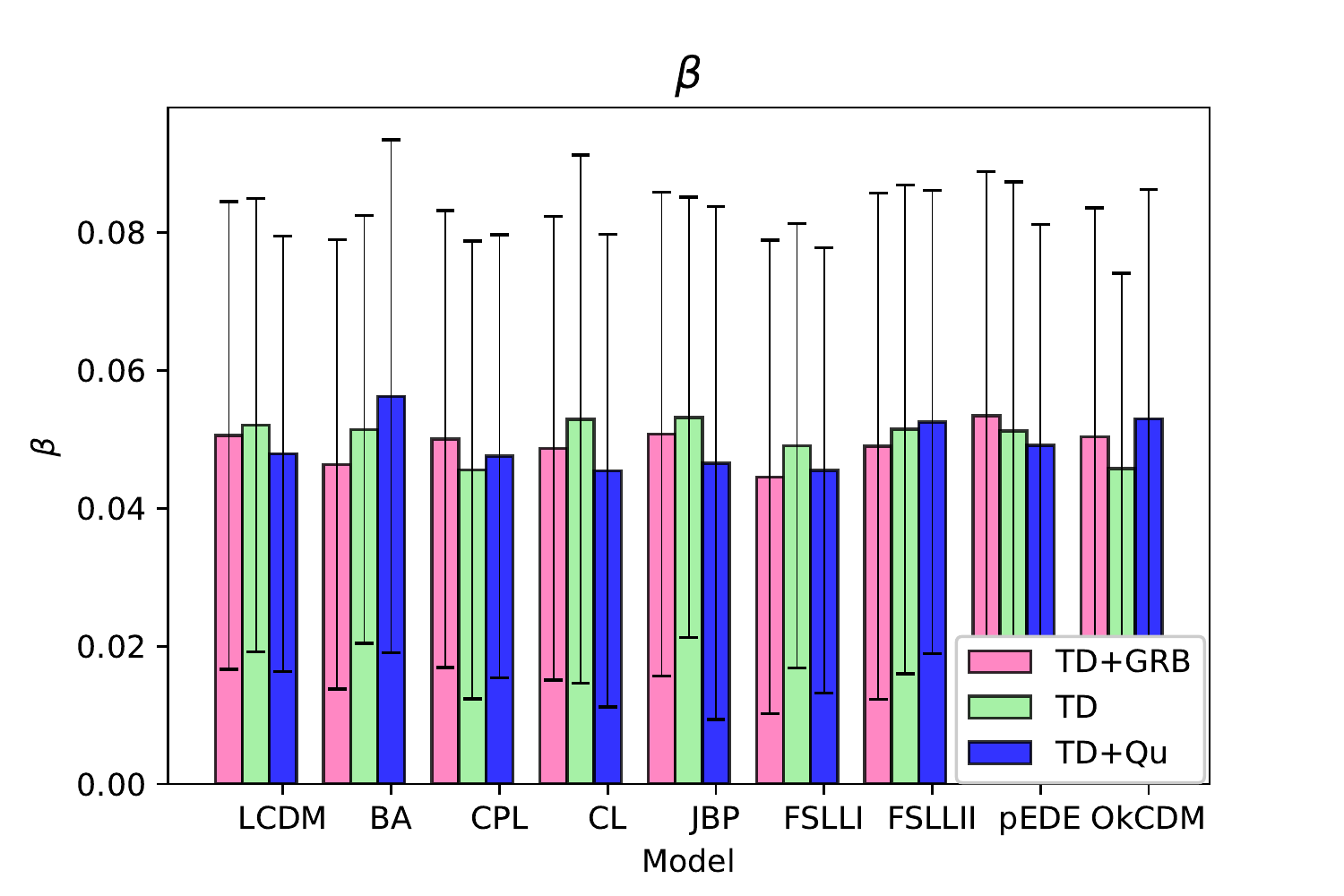}
\caption{The inferred values of the mean and error for the TD parameters $\alpha$ and $\beta$ for the second dataset, TD2}
\label{fig2}
\end{figure*}
\subsection{LIV parameters}
The results for TD1 and TD2 can be seen on Fig. \ref{fig1} and Fig. \ref{fig2} respectively. On them we present the time delay parameters $\alpha$ and $\beta$ depending on the cosmological model and on the used dataset: "TD", "TD+GRB" and "TD+Qu". On the plots we have displayed with bars the mean value in both cases and on top of it we have placed the error bars from the MCMC.

One can see that for TD1 in most cases the "TD+GRB" dataset gives similar results as the  "TD+Qu". The "TD" alone usually gives tighter constrains pointing that the two high-redshift datasets come with larger errors. The next important thing is that there is indeed a deviation between $\Lambda$CDM and the other models for the $\alpha$ parameter. The $\beta$ parameter on the contrary, looks much less variable and dependent on the cosmology, as expected, since our lack of knowledge on the intrinsic source is the same across the different models. The biggest deviation from $\Lambda$CDM is in $\Omega_K$CDM and JPB. 

For the second dataset, TD2, the errors are significantly larger for $\alpha$ while $\beta$ is about the same. We also see that this time "TD+GRB" dataset gives notably different results form the "TD+Qu" and "TD" datasets. This is especially visible in models such as BA, JPB and FSLLII. For TD1, for most models, the mean error on $\alpha$ within each model is larger than the deviation between different models, while for TD2, it can be smaller (for example $\Omega_k$CDM compared to the other models). 

If one wants to estimate the bounds on the $E_{QG}$, one would need to make an assumption about the value of $H_0$, since with this method, we do not infer it. We take the two most likely values of $H_0$: $H_0^{Planck}=67.4\pm0.5 \,{\rm km}\,{\rm s}^{-1}\,{\rm Mpc}^{-1}$ (the Planck 2018 CMB measurement \cite{Aghanim:2018eyx}) and $H_0^{R22}=73.04\pm1.04\,{\rm km}\,{\rm s}^{-1}\,{\rm Mpc}^{-1}$ (the local, model-independent measurement by SH0ES \cite{Riess:2021jrx}) and after converting it to $s^{-1}$ we obtain the following bounds for the energy scale, $E_{QG}$, for TD1: $E_{QG}^{Planck}\ge (5.4\pm 3.96)\times 10^{17}$ GeV and $E_{QG}^{SH0ES}\ge (4.98\pm 3.65)\times 10^{17}$GeV (for $\Lambda$CDM). For the second, TD2, dataset, we get as constraints: 
 $E_{QG}^{Planck}\ge (1.17\pm 0.89)\times 10^{17}$ GeV and $E_{QG}^{SH0ES}\ge (1.08\pm 0.83)\times 10^{17}$GeV (for $\Lambda$CDM). This is higher than the result in \cite{Vardanyan:2022ujc}, where the authors obtained $E_{QG}>4\times 10^{14}$GeV probably due to the inclusion of cosmology.  We summarize the results for the other models in Table 2. We see that the highest values for $E_{QG}$ are about $0.8\times 10^{17}$ GeV  higher than $\Lambda$CDM  for TD1 and $0.1\times 10^{17}$ GeV  for TD2. The lowest values are about $0.1\times 10^{17}$ GeV lower than $\Lambda$CDM for TD1 and $0.09\times 10^{17}$ GeV for TD2.
 
 Adding a curvature with the $\Omega_K$CDM model gives the lowest values of $\alpha$ and leads to deviation from $\Lambda$CDM: $E_{QG}^{TD1}\to E_{QG}^{\Lambda CDM} + 0.97\times 10^{17}$ GeV  and  $E_{QG}^{TD2}\to E_{QG}^{\Lambda CDM} + 0.6\times 10^{17}$ GeV respectively.
 
 We note that here we have taken the errors for both $H_0$ and $\alpha$ and $\Delta E_{TD1}=177$ keV and $\Delta E_{TD2}=100$ keV. Furthermore one needs to remember that the value of $b$ does not provide $H_0$ and the value itself is lower than expected, meaning if we take $r_d\approx 136$ Mpc, that will require $H_0\sim 78 \,{\rm km}\,{\rm s}^{-1}\,{\rm Mpc}^{-1}$. If we use this value, we get $E_{QG}>4.6\times10^{17}$ for TD1 and $E_{QG}>1.01\times10^{17}$ for TD2, i.e. slightly lower value, but still close to previous results. Measurements of $E_{QG}\ge 10^{17}$GeV have been already observed so they are not qualitatively new, but here we take the cosmological model into account, up to what the published data allows us.  For example, such bound of bound $E_{QG}>10^{17}$ GeV is obtained in \cite{Boggs:2003kxa} from GRB021206, in \cite{Xu:2016zsa} from GRB 160509A and \cite{Ellis:2018lca} from a collection of 8 GRBs and 5 different estimation procedures. Furthermore, there are plenty of works using model-independent approach, for example, \cite{Bezerra:2019vrz, Pan:2020zbl, Du:2020uev, Agrawal:2021cim,  Furtado:2021aod, Desai:2022hht} giving different bounds to $E_{QG}$. For example in \cite{Desai:2022hht}, the authors compare $\Lambda$CDM with respect to LIV and find negligible evidence for LIV  and $E_{QG} > 5\times 10^{15}$GeV.

In terms of variability of $\alpha$ depending on the cosmological model, we see that  deviation between the highest and the lowest result in TD1 is about 20\% in TD1 and it can reach 60\% in TD2. On the other hand, when translated to energy, due to the different $\Delta E$, one gets for $E_{QG}^{max}-E_{QG}^{min}$ for TD1 $\sim 1.1\times 10^{17}$GeV and for TD2$ \sim 0.69 \times 10^{17}$GeV for TD2. This means that while in terms of percents of the $\Lambda$CDM constraint, the deviation of TD2 is more than twice bigger, in absolute terms of energy, it is actually smaller. Since we do not know how the two datasets have been processed in terms of cosmology, it is hard to judge the origin of this variability and which estimate is better, in any case, that shows that the cosmology may account for at least 20\% deviation.  

\begin{table}
	\begin{center}
 \footnotesize
		\begin{tabular}{|c|c|c|c|c|}
			\hline
           {} & \multicolumn{2}{|c|}{$TD1\times10^{17}$GeV} & \multicolumn{2}{|c|}{$TD2\times10^{17}$GeV}\\ 
            \hline
			Model & $E^{Pl}_{QG}$ & $E^{SH0ES}_{QG}$ &  $E^{Pl}_{QG}$ & $E^{SH0ES}_{QG}$ \\
			\hline
			$\Lambda$CDM & $5.41\pm 4.00$ &  $4.98\pm 3.69$ & $1.17\pm 0.89$ & $1.08\pm 0.83$ \\
			\hline
			CPL & $5.31\pm 4.14$ & $4.88\pm 3.80$ & $1.08\pm 0.811$ & $0.99\pm 0.74$ \\
			\hline
			BA & $5.72\pm 4.35$ & $5.26\pm 4.00$ & $1.20\pm 0.9$ & $1.10\pm 0.82$  \\
			\hline
			LC & $5.64\pm 4.15$ & $5.19\pm 3.82$ & $1.04\pm 0.69$ & $0.96\pm 0.64$  \\
			\hline
			JPB & $6.25\pm 4.52$ & $5.75\pm 4.16$ & $1.38\pm 1.10$ & $1.27\pm 1.01$  \\
			\hline
			FSLLI & $6.06\pm 4.78$ & $5.57\pm 4.39$ & $1.14\pm 0.87$ & $1.04\pm 0.79$  \\
			\hline
			FSLLII & $5.72\pm 4.39$ & $5.26\pm 4.03$ & $1.19\pm 0.87$ & $1.09\pm 0.799$ \\
			\hline
			pEDE & $5.72\pm 4.15$ & $5.26\pm 3.81$ & $1.21\pm 0.90$ & $1.12\pm 0.84$ \\
			\hline
			OmegaK & $6.39\pm 4.78$ & $5.88\pm 4.40$ & $1.73\pm 1.33$ & $1.59\pm 1.22$ \\
			\hline
		\end{tabular}
	\end{center}
	\caption{Effect of the cosmological model on the bounds on $E_{QG}$.}
 \label{Table1}
\end{table}

\subsection{Cosmology}
\begin{figure*}
 	\centering
\includegraphics[width=0.49\textwidth]{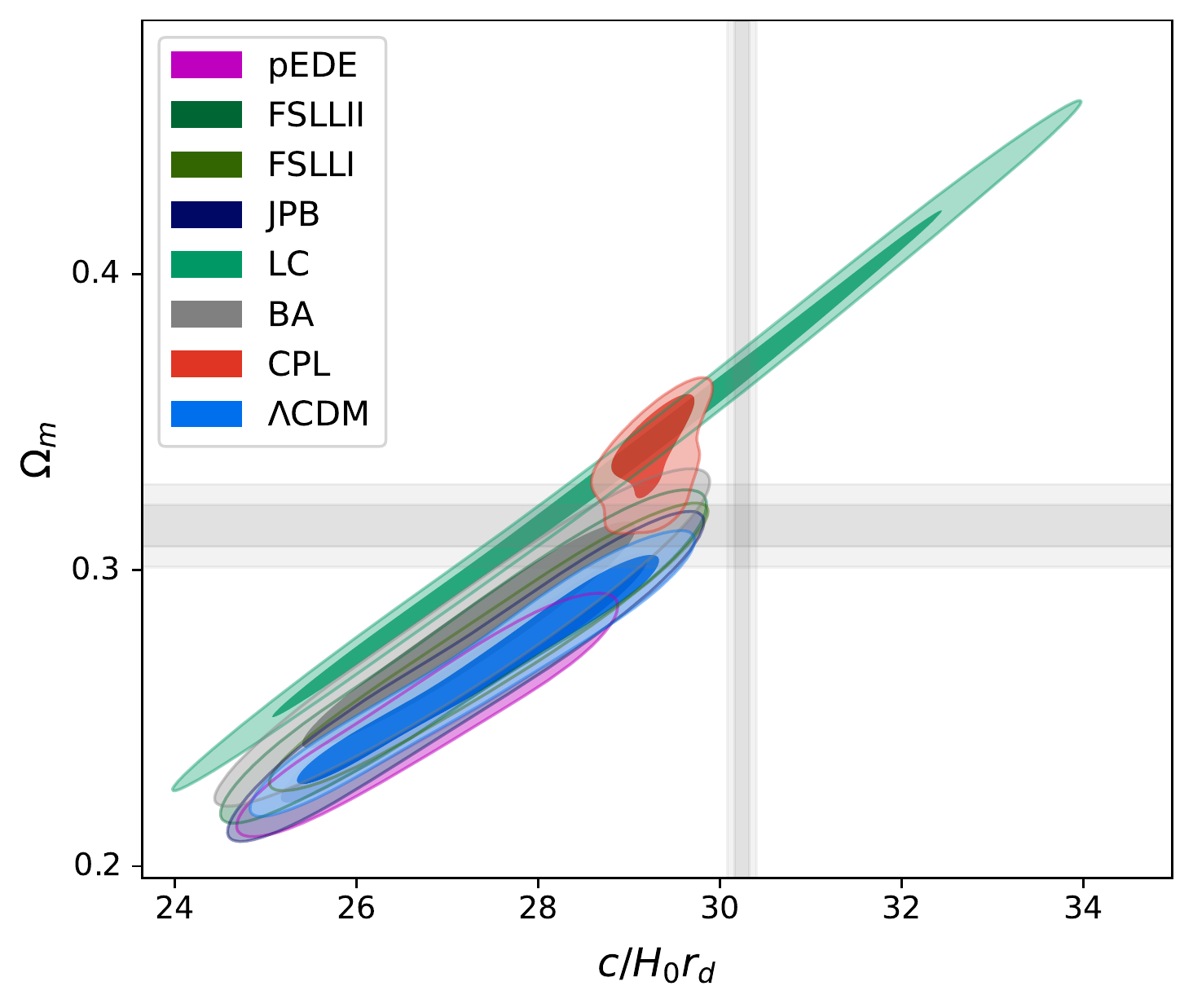}
\includegraphics[width=0.49\textwidth]{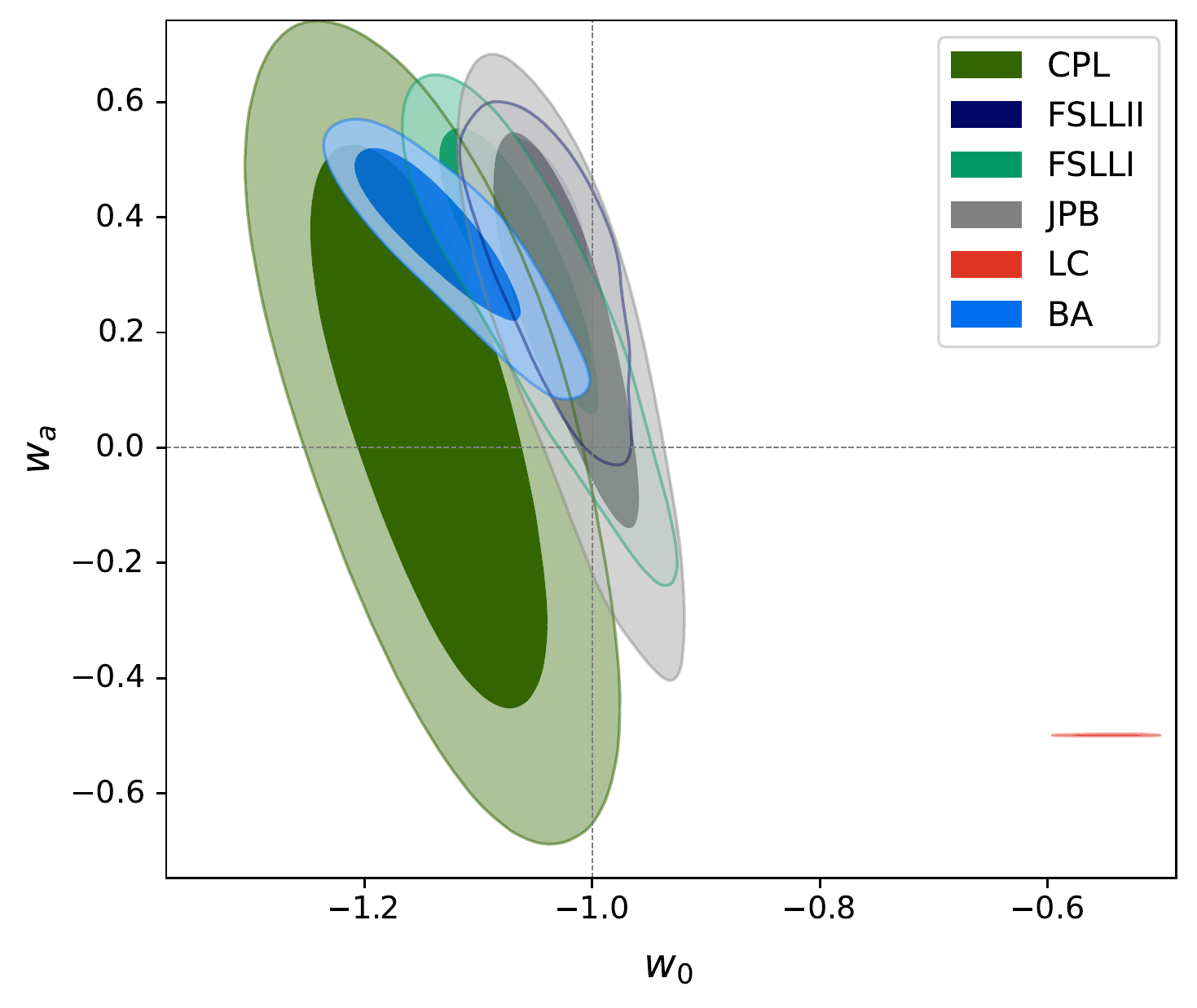}
\caption{The 2d posterior distribution at $68\% $ and $95\% $ CL for the cosmological parameters for TD1 $\Omega_m$ vs $c/(H_0 r_d)$ (left panel) and $w_a$ vs $w_0$ (right panel)}
\label{fig3}
\end{figure*}

\begin{figure*}
 	\centering
\includegraphics[width=0.49\textwidth]{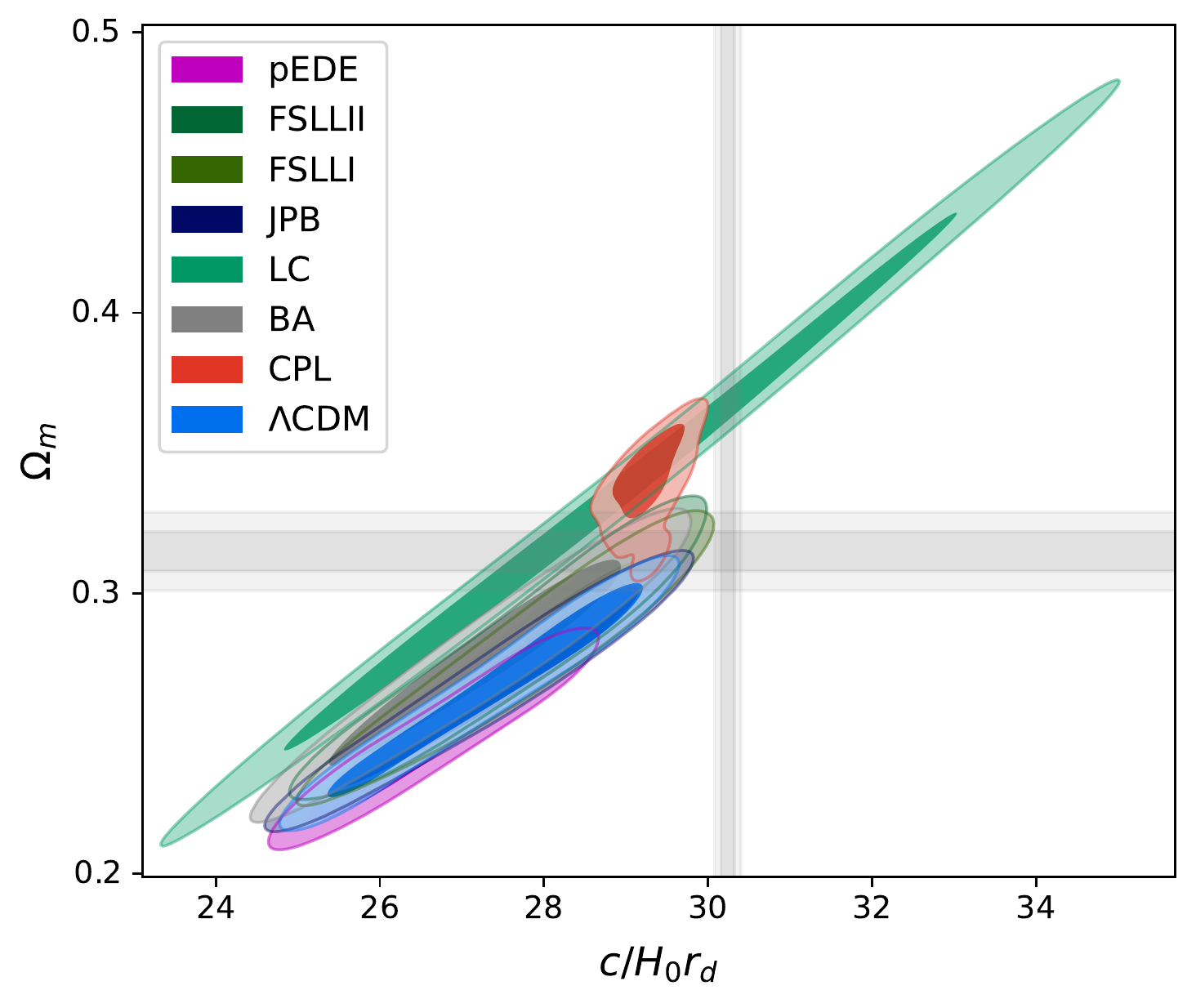}
\includegraphics[width=0.49\textwidth]{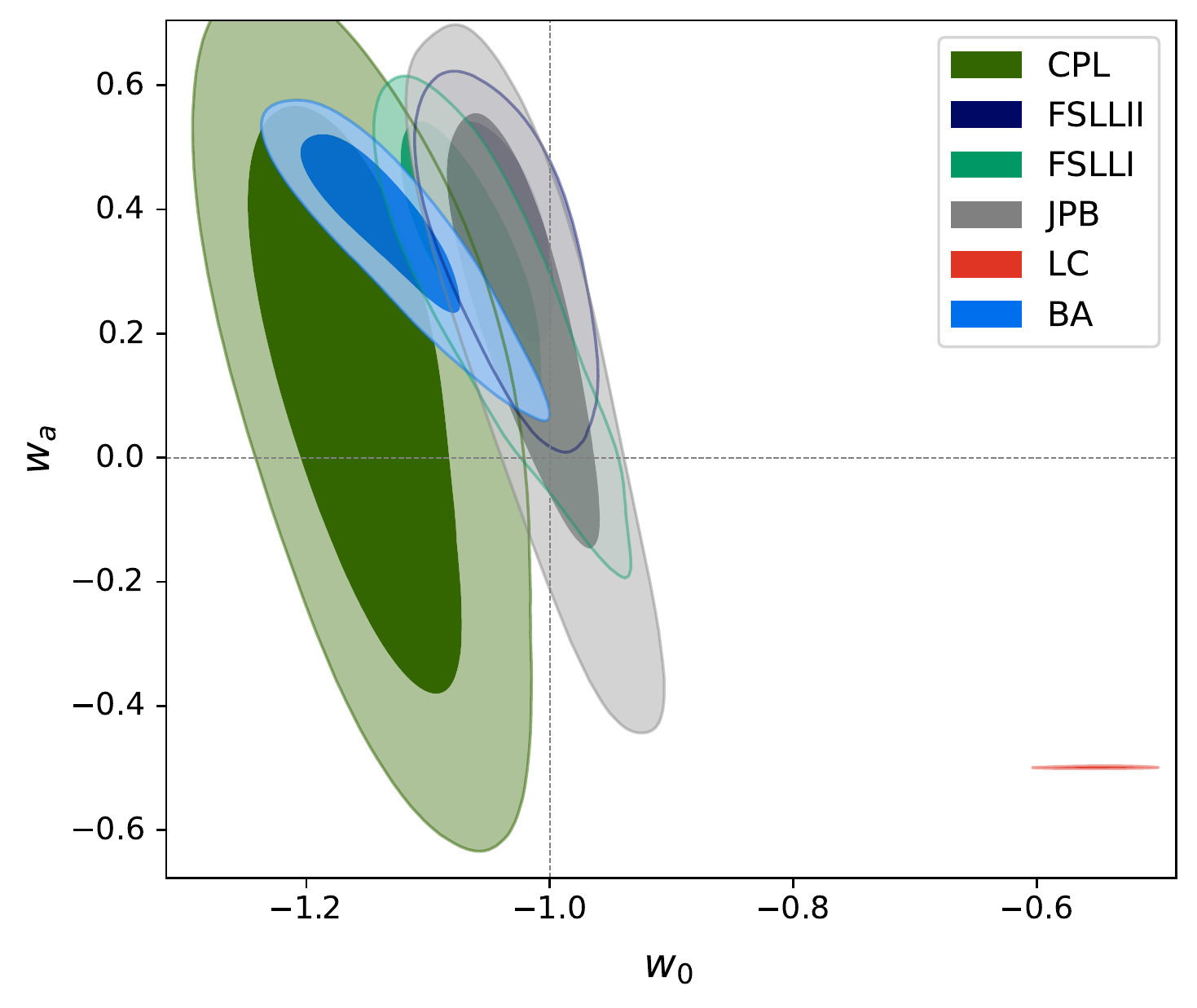}
\caption{The 2d posterior distribution at $68\% $ and $95\% $ CL for the cosmological parameters for TD2 $\Omega_m$ vs $c/(H_0 r_d)$ (left panel) and $w_a$ vs $w_0$ (right panel)}
\label{fig4}
\end{figure*}

The results for the cosmological parameters can be seen on Fig. \ref{fig3} and Fig. \ref{fig4}. Here we show the plots only for the TD dataset and do not plot the ones for TD+GRB and TD+Qua, since there is little change in the DE parameters from adding the additional datasets. We note that TD1 and TD2 yield very similar results showing that the cosmological study do not depend very strongly on the TD dataset. In all the examined cases we get a negative spatial curvature ($\Omega_K<0$). The constraints on all cosmological parameters from the two datasets can be found in the tables in the Appendix \ref{sec:appendix}.

An important result we obtained is that the inferred value for $b=c/H_0 r_d$ is lower than the value expected from the $\Lambda$CDM prediction-—for both Planck 2018 and SH0ES 2022 values of $H_0$ and $r_d$, one obtains $b\approx 30$. In the current work, we infer $b\sim 27.5$ in most cases. This differs from what we found in \cite{Staicova:2022zuh}, where we examined the same DE models without the TD datasets, and the result for this parameter was $b\sim 30$ for the $\Lambda$CDM model and $b\sim 28.5$ for the DE models. Peculiarly, in our previous work, adding the GRB dataset would lower the value of $b$. Here this effect is more pronounced, even without the said GRB dataset. It is important to note that while the TD dataset comes from GRBs, in it, one measures time delays, while in the GRB dataset, we measure the distance modulus. Thus, the two datasets are not equivalent. The value of $\Omega_m$, except for the CPL and LC models, is also consistently lower than the Planck 2018 value ($\Omega_m\sim 0.321 \pm 0.013$). Such a result has been already observed in some BAO studies and is curious, especially in the light of the Hubble tension and the degeneracy in the $H_0-r_d-\Omega_m$ plane in BAO measurements (see \cite{Knox:2019rjx, Staicova:2021ajb} for more discussion on the topic). From Planck 2018, the ratio $r_*/r_d\approx 0.979$, which matches our results, except for the $\Omega_K$CDM case. As for the DE parameters, they all look very close to previous studies, except that $w_0$ seems to be a little bit closer to the $\Lambda$CDM prediction than the one obtained in \cite{Staicova:2022zuh}. Note that the errors on $w_a$ here are smaller than those in \cite{Staicova:2022zuh} because our priors here are smaller due to numerical problems arising from larger priors.

Finally we use statistical measures to compare the models and to see if the introduction of the TD datasets changes the preference for $\Lambda$CDM that we have observed before. The definitions of the AIC, BIC, DIC and BF and the Jeffrey's scale can be found in \cite{Liddle:2007fy}. What is important is that, according to the definition we use (same as in \cite{Staicova:2022zuh}) in order for a model to compete with $\Lambda$CDM for IC measures, it has to have a positive $\Delta IC$ (and $>2$ for substantial support) and in terms of BF -- a negative one (and $<-1$ for substantial  support).  These results can be found in Table \ref{Table1}. From it, one can see that the AIC and BIC measures strongly prefer the $\Lambda$CDM model in both datasets TD1 and TD2. When it comes to the other two measures, we see that for the TD1 dataset, there is a small preference for DE models, specially CPL, BA and FSLLII. In terms of BF, the only contestant is BA. When it comes to the TD2 dataset, there is a preference again for CPL, BA and FSLLII in terms of DIC, but there is no preference for DE models in terms of BF, though BA is somewhat in the uncertainty zone. Again we see differences with \cite{Staicova:2022zuh} where most DE models were somewhat supported by DIC and BF. The full tables for all the datasets can be found in the Appendix. 

\begin{table}
	\begin{center}
 \footnotesize
		\begin{tabular}{|c|c|c|c|c|c|c|c|c|}
			\hline
            & \multicolumn{4}{|c|}{$TD1$} & \multicolumn{4}{|c|}{$TD2$}\\ 
            \hline
			Model & $\Delta$AIC & $\Delta BIC$ & $\Delta$DIC & ln(BF) & $\Delta$AIC & $\Delta BIC$ & $\Delta$DIC & ln(BF)\\
			\hline
			$\Lambda$CDM &  &  &  &   &  &  &  &\\
			\hline
			CPL & $-0.73$ & $-5.8$ & $3.1$ & $4.46$ &$-0.56$ & $-5.9$ & $3.3$ & $3.86$ \\
			\hline
			BA & $-0.92$ & $-6.0$ & $2.94$ & $-0.44$  & $-0.84$ & $-6.2$ & $2.96$ & $0.9$ \\
			\hline
			LC & $-208.6$ & $-213.7$ & $-204.8$ & $214.1$  & $-209.1$ & $-214.5$ & $-205.5$ & $214.2$\\
			\hline
			JPB & $-3.84$ & $-8.9$ & $0.073$ & $1.88$  & $-3.85$ & $-9.2$ & $0.12$ & $2.38$\\
			\hline
			FSLLI & $-2.82$ & $-7.9$ & $0.99$ & $1.88$ & $-3.1$ & $-8.4$ & $0.74$ & $2.02$  \\
			\hline
			FSLLII & $-2.06$ & $-7.1$ & $1.73$ & $1.56$  & $-1.99$ & $-7.3$ & $1.78$ & $2.03$\\
			\hline
			pEDE & $-22.7$ & $-22.7$ & $-22.7$ & $22.0$ & $-22.6$ & $-22.6$ & $-22.6$ & $23.5$ \\
			\hline
			OmegaK & $-58.1$ & $-60.6$ & $-56.4$ & $64.4$  & $-56.3$ & $-59.0$ & $-54.1$ & $63.5$\\
			\hline
		\end{tabular}
	\end{center}
	\caption{Excerpt from the statistical comparison between the different models, the full table can be found in the appendix. Everywhere we compare $\Lambda$CDM with the other cosmological models. TD1  \cite{Ellis:2005sjy} and and TD2 \cite{Vardanyan:2022ujc}.}
 \label{Table1}
\end{table}

\section{Discussion}
\label{sec:discussion}
The search for evidence of Lorentz Invariance Violation (LIV) has become a crucial aspect of the quest for a theory of quantum gravity. The discovery of GRBs with very high energy emission led to a significant advance in the field, yet the role of the cosmology hasn't been extensively discussed. In this article, we examine the dependence on the underlying cosmology of TD estimations based on GRB observations by studying 9 different cosmological models. Our additional datasets consist of a selection of BAO points, supplemented with SN points from the Pantheon dataset, and additional GRB dataset and a quasar dataset to account for high redshifts. We use a nested sampler to constrain the different cosmological and LIV parameters and to avoid setting a prior on $H_0$ we use the combined parameter $c/(H_0 r_d)$. Our results yield a limit of $E_{QG}>5\times 10^{17}$ GeV for TD1 and $E_{QG}>1.1\times 10^{17}$ GeV for TD2. Notably, we see that TD2 has bigger dependence on the cosmological model than TD1, with the deviation between the highest and lowest results being around 20\% for TD1 and reaching up to 60\% for TD2. This shows that while the cosmological model is not crucial for these TD datasets, it can amount to at least 20\%  of the estimate and thus it is important to be taken into account.  

A surprising side-effect of our study is that we obtain a lower value for $c/(H_0 r_d)$ compared to that of the BAO+GRBs sample alone (i.e. without accounting for time delays). This discrepancy is important since the deviation is $>1.8\sigma$, much larger than what was previously observed. This suggests that time-delay datasets could be a valuable addition to the cosmological studies toolbox, especially regarding solving the problem of the tensions. The error on this parameter is also larger than previously observed, indicating that the additional parameters are increasing the uncertainty of our cosmological results. 

In conclusion, we see that understanding the effect of the cosmological model on the LIV parameters is complex, particularly considering the uncertainties in the GRB progenitor's models and the tension in cosmological observations, but important. For future studies, full transparency regarding data processing is crucial, and efforts should be made to incorporate as much model-independent data as possible with respect to cosmology. This approach will facilitate future reference and inclusion in future datasets.
\ack
 D.S. is thankful to Bulgarian National Science Fund for support via research grant KP-06-N58/5. 

\bibliographystyle{JHEP}
\bibliography{ref1}

\appendix
\label{sec:appendix}
\newpage
\section{Tables with the cosmological parameters}

\begin{table}[!h]
	\begin{center}
 \scriptsize
\begin{tabular}{|c|c|c|c|c|c|c|c|}
			\hline
             \multicolumn{8}{|c|}{$TD+GRB$} \\
             \hline
			Model & $c/(H_0 r_d)$ & $\Omega_m$ & $r_s/r_d$ & w & $w_a$ &$\alpha \times 10^4$ & $\beta \times 10^2$\\
			\hline
   \multicolumn{8}{c|}{$TD+GRB$} \\
             \hline
			$\Lambda$CDM & $27.5\pm 1.5$ & $0.27\pm 0.03$ & $0.97\pm 0.01$ & -1.000 & 0.000 & $1.46\pm 1.11$ & $4.81\pm 3.4$ \\
			\hline
			CPL & $29.3\pm 0.2$ & $0.34\pm 0.01$ & $0.91\pm 0.01$ & $-1.14\pm 0.06$ & $0.11\pm 0.29$ & $1.66\pm 1.23$ & $4.88\pm 3.21$ \\
			\hline
			BA & $27.3\pm 1.6$ & $0.28\pm 0.03$ & $0.94\pm 0.01$ & $-1.13\pm 0.05$ & $0.37\pm 0.11$ & $1.41\pm 0.97$ & $4.91\pm 3.34$ \\
			\hline
			LC & $30.8\pm 1.9$ & $0.38\pm 0.05$ & $0.9\pm 0.0$ & $-0.55\pm 0.02$ & $-0.5\pm 0.0$ & $1.56\pm 1.17$ & $4.81\pm 3.63$ \\
			\hline
			JPB & $27.5\pm 1.5$ & $0.27\pm 0.03$ & $0.97\pm 0.01$ & $-1.02\pm 0.04$ & $0.23\pm 0.25$ & $1.43\pm 1.08$ & $4.42\pm 3.18$ \\
			\hline
			FSLLI & $27.4\pm 1.4$ & $0.27\pm 0.03$ & $0.96\pm 0.01$ & $-1.07\pm 0.04$ & $0.33\pm 0.12$ & $1.46\pm 1.01$ & $4.81\pm 3.63$ \\
			\hline
			FSLLII & $27.4\pm 1.5$ & $0.27\pm 0.03$ & $0.96\pm 0.01$ & $-1.04\pm 0.03$ & $0.35\pm 0.13$ & $1.43\pm 1.11$ & $4.71\pm 3.61$ \\
			\hline
			pEDE & $26.9\pm 1.2$ & $0.25\pm 0.02$ & $0.98\pm 0.01$ & - & - & $1.42\pm 1.03$ & $4.87\pm 3.22$ \\
			\hline
			$\Omega_K$CDM & $25.1\pm 0.1$ & $0.19\pm 0.0$ & $1.05\pm 0.01$ & $-6.87\pm 0.34^*$ & - & $1.37\pm 0.96$ & $5.13\pm 3.53$ \\
			\hline
		\end{tabular}
  
  \begin{tabular}{|c|c|c|c|c|c|c|c|}
	\hline
             \multicolumn{8}{|c|}{$TD$} \\		
   \hline
			\hline
			$\Lambda$CDM & $27.5\pm 1.5$ & $0.27\pm 0.03$ & $0.97\pm 0.01$ & -1.000 & 0.000 & $1.5\pm 1.11$ & $5.05\pm 3.46$ \\
			\hline
			CPL & $29.3\pm 0.2$ & $0.34\pm 0.01$ & $0.91\pm 0.01$ & $-1.14\pm 0.07$ & $4.56\pm 36.41$ & $1.53\pm 1.19$ & $5.22\pm 3.26$ \\
			\hline
			BA & $27.3\pm 1.2$ & $0.28\pm 0.03$ & $0.94\pm 0.01$ & $-1.13\pm 0.05$ & $0.37\pm 0.1$ & $1.42\pm 1.08$ & $4.66\pm 3.15$ \\
			\hline
			LC & $28.9\pm 2.7$ & $0.34\pm 0.06$ & $0.9\pm 0.0$ & $-0.55\pm 0.02$ & $-0.5\pm 0.0$ & $1.44\pm 1.06$ & $4.15\pm 2.61$ \\
			\hline
			JPB & $27.2\pm 1.5$ & $0.26\pm 0.03$ & $0.97\pm 0.01$ & $-1.02\pm 0.04$ & $0.21\pm 0.22$ & $1.3\pm 0.94$ & $4.77\pm 3.32$ \\
			\hline
			FSLLI & $27.7\pm 1.1$ & $0.28\pm 0.02$ & $0.96\pm 0.01$ & $-1.06\pm 0.05$ & $0.3\pm 0.16$ & $1.34\pm 1.06$ & $4.86\pm 3.04$ \\
			\hline
			FSLLII & $27.4\pm 1.5$ & $0.28\pm 0.03$ & $0.95\pm 0.01$ & $-1.04\pm 0.03$ & $0.35\pm 0.15$ & $1.42\pm 1.09$ & $4.73\pm 3.44$ \\
			\hline
			pEDE & $27.0\pm 1.2$ & $0.25\pm 0.02$ & $0.98\pm 0.01$ & - & - & $1.42\pm 1.03$ & $4.92\pm 3.21$ \\
			\hline
			$\Omega_K$CDM & $25.1\pm 0.0$ & $0.19\pm 0.0$ & $1.05\pm 0.01$ & $-6.86\pm 0.3^*$ & - & $1.27\pm 0.95$ & $5.0\pm 3.67$ \\
			\hline
		\end{tabular}
  
  \begin{tabular}{|c|c|c|c|c|c|c|c|}
	\hline
             \multicolumn{8}{|c|}{$TD+Qu$} \\
   \hline
   
			\hline
			$\Lambda$CDM & $27.5\pm 1.5$ & $0.27\pm 0.03$ & $0.97\pm 0.01$ & -1.000 & 0.000 & $1.43\pm 1.09$ & $4.89\pm 3.31$ \\
			\hline
			CPL & $29.2\pm 0.2$ & $0.34\pm 0.01$ & $0.92\pm 0.01$ & $-1.14\pm 0.06$ & $3.23\pm 33.64$ & $1.36\pm 1.04$ & $4.75\pm 3.17$ \\
			\hline
			BA & $27.7\pm 1.2$ & $0.28\pm 0.03$ & $0.95\pm 0.01$ & $-1.12\pm 0.06$ & $0.33\pm 0.13$ & $1.4\pm 0.96$ & $4.92\pm 3.28$ \\
			\hline
			LC & $30.4\pm 2.4$ & $0.37\pm 0.06$ & $0.9\pm 0.0$ & $-0.54\pm 0.02$ & $-0.5\pm 0.0$ & $1.65\pm 1.18$ & $4.84\pm 3.26$ \\
			\hline
			JPB & $27.3\pm 1.6$ & $0.26\pm 0.03$ & $0.97\pm 0.01$ & $-1.02\pm 0.04$ & $0.18\pm 0.28$ & $1.38\pm 1.01$ & $5.02\pm 3.48$ \\
			\hline
			FSLLI & $27.7\pm 1.2$ & $0.28\pm 0.03$ & $0.96\pm 0.01$ & $-1.06\pm 0.04$ & $0.29\pm 0.17$ & $1.46\pm 1.12$ & $5.07\pm 3.51$ \\
			\hline
			FSLLII & $27.3\pm 1.1$ & $0.27\pm 0.02$ & $0.96\pm 0.01$ & $-1.04\pm 0.03$ & $0.32\pm 0.13$ & $1.4\pm 0.99$ & $4.82\pm 3.42$ \\
			\hline
			pEDE & $26.9\pm 1.1$ & $0.25\pm 0.02$ & $0.98\pm 0.01$ & - & - & $1.27\pm 0.92$ & $4.72\pm 3.51$ \\
			\hline
			$\Omega_K$CDM & $25.4\pm 0.3$ & $0.21\pm 0.01$ & $1.01\pm 0.01$ & $-3.29\pm 0.32^*$ & - & $1.31\pm 0.97$ & $5.01\pm 3.44$ \\
			\hline
		\end{tabular}
 
	\end{center}
	\caption{Final values of the inferred parameters for the considered models from the TD1 dataset. The value denoted with ${}^*$ corresponds to $\Omega_K \times  10^2$}
 \label{tab:allTD1}
\end{table}

\begin{table}[!h]
	\begin{center}
 \scriptsize
\begin{tabular}{|c|c|c|c|c|c|c|c|}
			\hline
             \multicolumn{8}{|c|}{$TD+GRB$} \\
             \hline
			Model & $c/(H_0 r_d)$ & $\Omega_m$ & $r_s/r_d$ & w & $w_a$ & $\alpha \times 10^4$ & $\beta \times 10^2$ \\
			\hline
			\hline
			$\Lambda$CDM & $27.6\pm 1.4$ & $0.27\pm 0.03$ & $0.97\pm 0.01$ & -1.000 & 0.000 & $4.07\pm 2.98$ & $5.06\pm 3.39$ \\
			\hline
			CPL & $29.3\pm 0.2$ & $0.34\pm 0.01$ & $0.92\pm 0.01$ & $-1.14\pm 0.06$ & None & $4.43\pm 3.48$ & $4.64\pm 3.26$ \\
			\hline
			BA & $27.4\pm 1.3$ & $0.28\pm 0.03$ & $0.94\pm 0.01$ & $-1.13\pm 0.04$ & $0.38\pm 0.1$ & $4.26\pm 3.27$ & $5.01\pm 3.32$ \\
			\hline
			LC & $31.5\pm 1.5$ & $0.4\pm 0.04$ & $0.9\pm 0.0$ & $-0.55\pm 0.02$ & $-0.5\pm 0.0$ & $4.23\pm 3.2$ & $4.87\pm 3.36$ \\
			\hline
			JPB & $27.8\pm 1.3$ & $0.28\pm 0.03$ & $0.96\pm 0.01$ & $-1.02\pm 0.04$ & $0.25\pm 0.2$ & $4.17\pm 3.12$ & $5.08\pm 3.51$ \\
			\hline
			FSLLI & $27.4\pm 1.6$ & $0.27\pm 0.03$ & $0.96\pm 0.01$ & $-1.07\pm 0.04$ & $0.35\pm 0.12$ & $3.65\pm 2.88$ & $4.46\pm 3.44$ \\
			\hline
			FSLLII & $27.6\pm 1.5$ & $0.28\pm 0.03$ & $0.96\pm 0.01$ & $-1.04\pm 0.03$ & $0.36\pm 0.12$ & $4.31\pm 3.1$ & $4.9\pm 3.67$ \\
			\hline
			pEDE & $27.0\pm 1.2$ & $0.25\pm 0.02$ & $0.98\pm 0.01$ & - & - & $4.43\pm 3.33$ & $5.34\pm 3.55$ \\
			\hline
			$\Omega_K$CDM & $25.2\pm 0.1$ & $0.2\pm 0.0$ & $1.02\pm 0.01$ & $-4.58\pm 0.27^*$ & - & $3.68\pm 2.77$ & $5.04\pm 3.32$ \\
			\hline
			\end{tabular}
			\begin{tabular}{|c|c|c|c|c|c|c|c|}
   \multicolumn{8}{c|}{$TD$} \\
             \hline
			$\Lambda$CDM & $27.4\pm 1.5$ & $0.27\pm 0.03$ & $0.97\pm 0.01$ & -1.000 & 0.000 & $3.91\pm 2.99$ & $5.21\pm 3.29$ \\
			\hline
			CPL & $29.3\pm 0.2$ & $0.34\pm 0.01$ & $0.91\pm 0.01$ & $-1.15\pm 0.06$ & $0.11\pm 0.34$ & $4.26\pm 3.19$ & $5.14\pm 3.11$ \\
			\hline
			BA & $27.3\pm 1.2$ & $0.28\pm 0.03$ & $0.95\pm 0.01$ & $-1.13\pm 0.04$ & $0.38\pm 0.1$ & $3.83\pm 2.86$ & $4.56\pm 3.32$ \\
			\hline
			LC & $28.8\pm 2.9$ & $0.34\pm 0.07$ & $0.9\pm 0.0$ & $-0.55\pm 0.02$ & $-0.5\pm 0.0$ & $4.41\pm 2.94$ & $5.29\pm 3.83$ \\
			\hline
			JPB & $27.4\pm 1.4$ & $0.27\pm 0.03$ & $0.97\pm 0.01$ & $-1.02\pm 0.04$ & $0.2\pm 0.24$ & $3.32\pm 2.65$ & $5.32\pm 3.19$ \\
			\hline
			FSLLI & $27.8\pm 1.1$ & $0.28\pm 0.02$ & $0.96\pm 0.01$ & $-1.06\pm 0.04$ & $0.31\pm 0.16$ & $4.04\pm 3.07$ & $4.91\pm 3.22$ \\
			\hline
			FSLLII & $27.7\pm 1.4$ & $0.28\pm 0.03$ & $0.95\pm 0.01$ & $-1.04\pm 0.03$ & $0.36\pm 0.12$ & $3.86\pm 2.83$ & $5.15\pm 3.55$ \\
			\hline
			pEDE & $26.6\pm 1.1$ & $0.25\pm 0.02$ & $0.98\pm 0.01$ & - & - & $3.78\pm 2.82$ & $5.12\pm 3.61$ \\
			\hline
			$\Omega_K$CDM & $25.0\pm 0.0$ & $0.19\pm 0.0$ & $1.04\pm 0.01$ & $-6.81\pm 0.22^*$ & - & $2.65\pm 2.03$ & $4.58\pm 2.83$ \\
			\hline
		\end{tabular}
  \begin{tabular}{|c|c|c|c|c|c|c|c|}
	\hline
             \multicolumn{8}{|c|}{$TD+Qu$} \\
   \hline$\Lambda$CDM & $27.7\pm 1.1$ & $0.27\pm 0.02$ & $0.97\pm 0.01$ & -1.000 & 0.000 & $4.18\pm 2.96$ & $4.79\pm 3.16$ \\
			\hline
			CPL & $29.2\pm 0.2$ & $0.33\pm 0.01$ & $0.92\pm 0.02$ & $-1.15\pm 0.05$ & $0.12\pm 0.36$ & $3.46\pm 2.63$ & $5.63\pm 3.72$ \\
			\hline
			BA & $27.3\pm 1.4$ & $0.28\pm 0.03$ & $0.95\pm 0.01$ & $-1.12\pm 0.05$ & $0.33\pm 0.14$ & $3.87\pm 2.79$ & $4.76\pm 3.22$ \\
			\hline
			LC & $28.7\pm 2.3$ & $0.33\pm 0.05$ & $0.9\pm 0.0$ & $-0.54\pm 0.02$ & $-0.5\pm 0.0$ & $4.11\pm 3.07$ & $4.55\pm 3.43$ \\
			\hline
			JPB & $27.3\pm 1.4$ & $0.27\pm 0.03$ & $0.97\pm 0.01$ & $-1.02\pm 0.04$ & $0.18\pm 0.23$ & $3.92\pm 2.72$ & $4.66\pm 3.72$ \\
			\hline
			FSLLI & $27.3\pm 1.5$ & $0.27\pm 0.03$ & $0.96\pm 0.01$ & $-1.07\pm 0.04$ & $0.26\pm 0.2$ & $3.88\pm 2.82$ & $4.55\pm 3.23$ \\
			\hline
			FSLLII & $27.6\pm 1.5$ & $0.28\pm 0.03$ & $0.96\pm 0.01$ & $-1.04\pm 0.03$ & $0.33\pm 0.13$ & $3.53\pm 2.72$ & $5.25\pm 3.36$ \\
			\hline
			pEDE & $26.7\pm 1.2$ & $0.25\pm 0.02$ & $0.98\pm 0.01$ & - & - & $3.67\pm 2.68$ & $4.91\pm 3.2$ \\
			\hline
			$\Omega_K$CDM & $25.1\pm 0.1$ & $0.19\pm 0.0$ & $1.05\pm 0.01$ & $-6.91\pm 0.34^*$ & - & $3.41\pm 2.65$ & $5.3\pm 3.33$ \\
			\hline

			\hline
		\end{tabular}
 
	\end{center}
	\caption{Final values of the inferred parameters for the considered models from the TD2 dataset. The value denoted with ${}^*$ corresponds to $\Omega_K\times 10^2$}
 \label{tab:allTD2}
\end{table}


\begin{table}[!ht]
\centering
\scriptsize
\begin{tabular}{ccc}
		\begin{tabular}{|c|c|c|c|c|}
			\hline
               &\multicolumn{4}{c|}{$TD+GRB$} \\
               \hline
			Model & $\Delta$AIC & $\Delta BIC$ & $\Delta$DIC & ln(BF) \\
			\hline
			$\Lambda$CDM &  &  &  &  \\
			\hline
			CPL & $-0.6$ & $-8$ & $3.2$ & $4$ \\
			\hline
			BA & $-0.4$ & $-7$ & $3.5$ & $-1$ \\
			\hline
			LC & $-209$ & $-216$ & $-206$ & $216$ \\
			\hline
			JPB & $-3.8$ & $-11$ & $-3.8$ & $2.2$ \\
			\hline
			FSLLI & $-2.5$ & $-10$ & $1.4$ & $1.1$ \\
			\hline
			FSLLII & $-1.9$ & $-9$ & $1.7$ & $1.2$ \\
			\hline
			pEDE & $-23$ & $-23$ & $-23$ & $24$ \\
			\hline
			OmegaK & $-59$ & $-63$ & $-58$ & $67$ \\
			\hline
		\end{tabular}

\begin{tabular}{|c|c|c|c|}
			\hline
   \multicolumn{4}{|c|}{$TD$} \\
               \hline
			$\Delta$AIC & $\Delta BIC$ & $\Delta$DIC & ln(BF) \\
			\hline
			 &  &  &  \\
			\hline
			$-0.7$ & $-6$ & $3.1$ & $4.5$ \\
			\hline
			$-0.9$ & $-6$ & $2.9$ & $-0.4$ \\
			\hline
			$-209$ & $-214$ & $-205$ & $214$ \\
			\hline
			$-3.8$ & $-9$ & $0.07$ & $1.9$ \\
			\hline
			$-2.8$ & $-8$ & $1$ & $1.9$ \\
			\hline
			$-2.1$ & $-7$ & $1.7$ & $1.6$ \\
			\hline
			$-23$ & $-23$ & $-23$ & $22$ \\
			\hline
			$-58$ & $-61$ & $-56$ & $64$ \\
			\hline
		\end{tabular}

     \begin{tabular}{|c|c|c|c|}
			\hline
   \multicolumn{4}{|c|}{$TD+Qu$} \\
               \hline
			$\Delta$AIC & $\Delta BIC$ & $\Delta$DIC & ln(BF) \\
			\hline
			 &  &  &  \\
			\hline
			$-1.5$ & $-7$ & $2.4$ & $5$ \\
			\hline
			$-2.2$ & $-8$ & $1.7$ & $3.7$ \\
			\hline
			$-220$ & $-225$ & $-216$ & $226$ \\
			\hline
			$-4.3$ & $-10$ & $-0.4$ & $2.4$ \\
			\hline
			$-3.4$ & $-9$ & $0.5$ & $1.9$ \\
			\hline
			$-2.7$ & $-8$ & $1.2$ & $1.3$ \\
			\hline
			$-21$ & $-21$ & $-20$ & $20$ \\
			\hline
			$-15$ & $-18$ & $-14$ & $18$ \\
			\hline
		\end{tabular}
\end{tabular}
\caption{Statistical measures for the TD1 dateset, where we rounded numbers to fit the table.}
\label{tab:statTD1}
\end{table}

\begin{table}[!ht]
\centering
\scriptsize
\begin{tabular}{ccc}
		\begin{tabular}{|c|c|c|c|c|}
			\hline
               &\multicolumn{4}{c|}{$TD+GRB$} \\
               \hline
			Model & $\Delta$AIC & $\Delta BIC$ & $\Delta$DIC & ln(BF) \\
			\hline
			$\Lambda$CDM &  &  &  &  \\
			\hline
			CPL & $-0.7$ & $-8$ & $3.1$ & $6$ \\
			\hline
			BA & $-0.4$ & $-8$ & $3.5$ & $-0.4$ \\
			\hline
			LC & $-209$ & $-216$ & $-205$ & $215$ \\
			\hline
			JPB & $-3.7$ & $-11$ & $0.2$ & $1.3$ \\
			\hline
			FSLLI & $-2.3$ & $-9$ & $1.7$ & $0.6$ \\
			\hline
			FSLLII & $-1.9$ & $-9$ & $1.9$ & $1.1$ \\
			\hline
			pEDE & $-23$ & $-23$ & $-23$ & $24$ \\
			\hline
			OmegaK & $-28$ & $-32$ & $-26$ & $34$ \\
			\hline
		\end{tabular}

\begin{tabular}{|c|c|c|c|}
			\hline
   \multicolumn{4}{|c|}{$TD$} \\
               \hline
			$\Delta$AIC & $\Delta BIC$ & $\Delta$DIC & ln(BF) \\
			\hline
			 &  &  &  \\
			\hline
			$-0.6$ & $-6$ & $3.3$ & $3.9$ \\
			\hline
			$-0.8$ & $-6$ & $3$ & $0.9$ \\
			\hline
			$-209$ & $-214$ & $-205$ & $214$ \\
			\hline
			$-3.9$ & $-9$ & $0.1$ & $2.4$ \\
			\hline
			$-3.1$ & $-8$ & $0.7$ & $2$ \\
			\hline
			$-2$ & $-7$ & $1.8$ & $2$ \\
			\hline
			$-23$ & $-23$ & $-23$ & $23$ \\
			\hline
			$-56$ & $-59$ & $-54$ & $64$ \\
			\hline
		\end{tabular}

     \begin{tabular}{|c|c|c|c|}
			\hline
   \multicolumn{4}{|c|}{$TD+Qu$} \\
               \hline
			$\Delta$AIC & $\Delta BIC$ & $\Delta$DIC & ln(BF) \\
			\hline
			 &  &  &  \\
			\hline
			$-1.1$ & $-7$ & $2.9$ & $6$ \\
			\hline
			$-2.1$ & $-8$ & $1.7$ & $1.8$ \\
			\hline
			$-219$ & $-225$ & $-215$ & $223$ \\
			\hline
			$-4.2$ & $-10$ & $-0.3$ & $0.7$ \\
			\hline
			$-3.6$ & $-9$ & $0.3$ & $2.2$ \\
			\hline
			$-2.8$ & $-9$ & $1$ & $1.1$ \\
			\hline
			$-20$ & $-20$ & $-20$ & $20$ \\
			\hline
			$-55$ & $-58$ & $-54$ & $61$ \\
			\hline
		\end{tabular}
\end{tabular}
\caption{Statistical measures for the TD2 dateset, where we rounded numbers to fit the table.}
\label{tab:statTD2}
\end{table}


\end{document}